\documentclass[a4paper,conference]{IEEEtran}

\usepackage{subcaption}
\usepackage{cite}
\usepackage{amsmath,amssymb,amsfonts}
\usepackage{algorithmic}
\usepackage{graphicx}
\usepackage{textcomp}
\usepackage{xcolor}
\usepackage[acronym]{glossaries}
\usepackage{hyperref}
\usepackage{rotating}
\usepackage{multirow}
\usepackage{graphicx}
\usepackage{url}

\DeclareMathOperator{\atantwo}{atan2}
\DeclareMathOperator{\relu}{ReLU}
\DeclareMathOperator{\LN}{LN}
\DeclareMathOperator{\sconv2d}{SeparableConv2D}
\DeclareMathOperator{\tconv2d}{Conv2DTranspose}
\DeclareMathOperator{\nconv2d}{Conv2D}
\DeclareMathOperator{\istft}{iSTFT}
\DeclareMathOperator{\stft}{STFT}
\DeclareMathOperator{\mha}{MHA}
\DeclareMathOperator{\drop}{Dropout}
\DeclareMathOperator{\ffw}{FFW}

\newacronym{FFW}{FFW}{feed-forward}
\newacronym{ES}{ES}{Electronic Support}
\newacronym{ESM}{ESM}{Electronic Support Measures}
\newacronym{LPI}{LPI}{Low Probability of Intercept}
\newacronym[plural = SDRs]{SDR}{SDR}{software defined radio}
\newacronym{PRI}{PRI}{Pulse Repetition Interval}
\newacronym{GPU}{GPU}{Graphics Processing Unit}
\newacronym{CW}{CW}{continuous wave}
\newacronym{EW}{EW}{Electronic Warfare}
\newacronym[plural = PDWs]{PDW}{PDW}{pulse descriptor word}
\newacronym[plural = EDWs]{EDW}{EDW}{emitter descriptor word}
\newacronym{TOA}{TOA}{time of arrival}
\newacronym{TDOA}{TDOA}{time difference of arrival}
\newacronym{AOA}{AOA}{angle of arrival}
\newacronym{RF}{RF}{radio frequency}
\newacronym{PW}{PW}{pulse width}
\newacronym{BW}{BW}{bandwidth}
\newacronym{PA}{PA}{pulse amplitude}
\newacronym[plural = RNNs]{RNN}{RNN}{recurrent neural network}
\newacronym{SNR}{SNR}{signal-to-noise ratio}
\newacronym[plural = STFTs]{STFT}{STFT}{short-time Fourier transform}
\newacronym{FFT}{FFT}{fast Fourier transform}
\newacronym{iFFT}{iFFT}{inverse fast Fourier transform}
\newacronym{DP-TF-Transformer}{DP-TF-Transformer}{Dual-Path Time-Frequency Transformer}
\newacronym[plural = iSTFTs]{iSTFT}{iSTFT}{inverse short-time Fourier transform}
\newacronym{LN}{LN}{Layer Normalization}
\newacronym{ReLU}{ReLU}{Rectified Linear Unit}
\newacronym[plural = F-TEs]{F-TE}{F-TE}{Frequency Transformer Encoder}
\newacronym[plural = T-TEs]{T-TE}{T-TE}{Time Transformer Encoder}
\newacronym{MHA}{MHA}{Multi-Head Attention}
\newacronym{SD-SDR}{SD-SDR}{scale-dependent signal-to-distortion ratio}
\newacronym{uPIT}{uPIT}{utterance-level Permutation Invariant Training}
\newacronym{PCA}{PCA}{Principal Component Analysis}
\newacronym{ICA}{ICA}{Independent Component Analysis}
\newacronym{NMF}{NMF}{Non-negative matrix factorization}
\newacronym{SIGINT}{SIGINT}{Signals Intelligence}

\glsaddall

\def\BibTeX{{\rm B\kern-.05em{\sc i\kern-.025em b}\kern-.08em
		T\kern-.1667em\lower.7ex\hbox{E}\kern-.125emX}}
	
\begin{document}
\title{Blind Source Separation of Radar Signals in Time Domain Using Deep Learning\\
}
\captionsetup[table]{position=bottom}
\renewcommand{\arraystretch}{1.2}

\author{\IEEEauthorblockN{Sven Hinderer\textsuperscript{*$\dag$}}
\IEEEauthorblockA{
\textit{\textsuperscript{*}Fraunhofer Institute for High Frequency Physics and Radar Techniques FHR, 53343 Wachtberg, Germany}\\
\textit{\textsuperscript{$\dag$}Institute of Signal Processing and System Theory, University of Stuttgart, 70569 Stuttgart, Germany}
 \\
sven.hinderer@fhr.fraunhofer.de\\ sven.hinderer@iss.uni-stuttgart.de}}

\maketitle

\noindent\textbf{Note:} This is the accepted version of the paper. 
The final version is published in the \emph{Proceedings of the IEEE 2022 23rd International Radar Symposium (IRS)}. 
DOI: \href{https://doi.org/10.23919/IRS54158.2022.9904990}{10.23919/IRS54158.2022.9904990}

\vspace{1em}

\renewcommand{\thefootnote}{}
\footnotetext{
	\textcopyright~2022 Personal use of this material is permitted.  Permission from IEEE must be obtained for all other uses, in any current or future media, including reprinting/republishing this material for advertising or promotional purposes, creating new collective works, for resale or redistribution to servers or lists, or reuse of any copyrighted component of this work in other works
}
\renewcommand{\thefootnote}{\arabic{footnote}}

\begin{abstract}
Identification and further analysis of radar emitters in a contested environment requires detection and separation of incoming signals. If they arrive from the same direction and at similar frequencies, deinterleaving them remains challenging. A solution to overcome this limitation becomes increasingly important with the advancement of emitter capabilities. We propose treating the problem as blind source separation in time domain and apply supervisedly trained neural networks to extract the underlying signals from the received mixture. This allows us to handle highly overlapping and also~\gls{CW} signals from both radar and communication emitters. We make use of advancements in the field of audio source separation and extend a current state-of-the-art model with the objective of deinterleaving arbitrary~\gls{RF} signals. Results show, that our approach is capable of separating two unknown waveforms in a given frequency band with a single channel receiver.
\end{abstract}


\section{Introduction}
An important task in the field of passive surveillance is the analysis of hostile radar emitters to estimate their mode of operation, their potential threat level, and to gain situational awareness. In a modern~\gls{EW} environment, we usually receive multiple signals from different radio emitters simultaneously. It is therefore required to separate these signals and assign them to their emitters before applying further signal processing operations.\newline
As of now, this problem is solved by first separating radar and communication signals. Radar signals are further divided into pulsed and \gls{CW} signals. Pulsed radar signals are described by~\glspl{PDW} and the extracted radar signal parameters are used to represent emitters with~\glspl{EDW}. Pulse parameters to describe radar signals are e.g. the~\gls{TOA},~\gls{PW},~\gls{AOA},~\gls{PA} and~\gls{RF}. A comparison of various popular approaches to deinterleave the extracted train of~\glspl{PDW} can be found in~\cite{comp_pri_deinter}. Newer litarature solves the problem with self-attention based soft min-cost flow learning, showing promising results~\cite{attn_flow_pri}.\newline
Although these deinterleaving approaches work well and are able to separate pulse trains under difficult conditions, they might fall short in separating signals in more complex scenarios. These include separation of not only pulsed, but also~\gls{CW} signals from both radar and communication emitters with a single channel receiver or if they originate from the same direction, dealing with pulses widely spread in frequency and time and also detecting and separating them under low~\gls{SNR} conditions. \newline
We propose treating the problem as blind source separation and use deep neural networks to separate the raw baseband input mixture in time domain. Blind source separation is an ongoing research topic, where huge advancements have been made over the last years by using deep neural networks~\cite{deep_clustering, convtasnet, dprnn, trans_speech}. Since this approach makes no assumptions about the inputs besides them being in the frequency band of interest, it is able to handle the aforementioned shortcomings of current methods. It comes, however, at the cost of high computational complexity and the need for extensive training data to achieve good separation performance. In this first study, we restrict us to a single input channel and two superposed signals. We adapt ideas from state-of-the-art neural network architectures in the field of single channel audio source separation to separate~\gls{RF} signals that can overlap in frequency and time. This is done by estimating two filter masks, which divide the mixed input into two separate components. 
\section{Model}
\subsection{Overview}
\begin{figure}
	\centering{
		\includegraphics[width=1.0\linewidth]{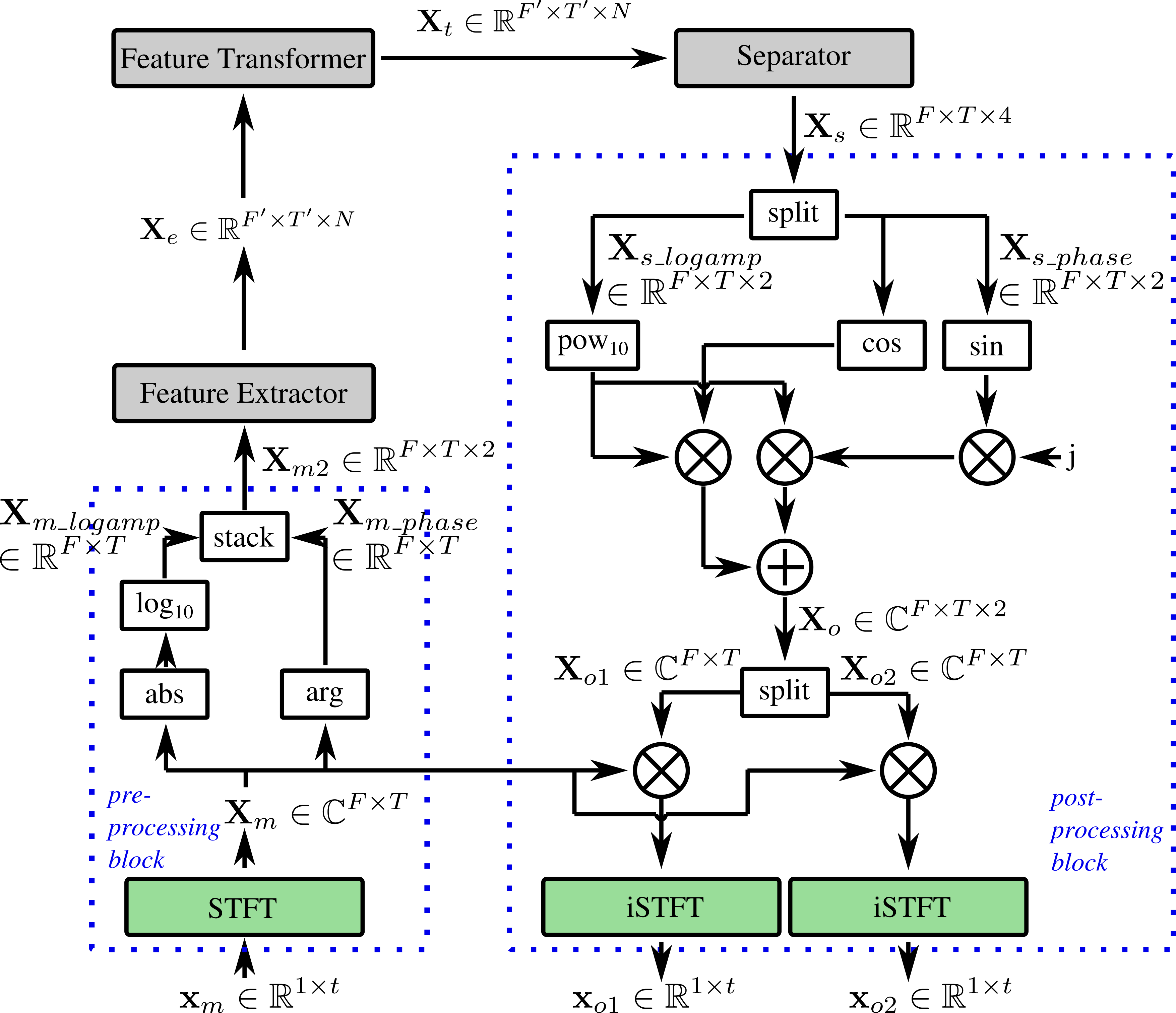}}
	\caption{Overview of the whole system: In the preprocessing, we convert the superposition of signals from time domain to time-frequency domain and apply transformations to handle the dynamic range of amplitudes. The processed input is then fed to our three trainable network modules. The predicted masks from the Separator are transformed again in the postprocessing stage, multiplied with the input mixture in time-frequency domain and the resulting separated signals are converted back to time domain.}
	\label{fig:model_overview}
\end{figure}

 An overview of the~\gls{RF} signal separation system is shown in Fig.~\ref{fig:model_overview}. To handle the bandwidth of received signals and their amplitudes' dynamic range, which are much larger for~\gls{RF} signals than what is investigated in the audio domain, multiple novel elements are added to the processing chain. These include differentiable~\glspl{STFT} and~\glspl{iSTFT} instead of 1D convolutional layers, since they are very fast by applying the~\gls{FFT} and~\gls{iFFT} algorithms. They are also strictly information preserving with appropriate settings, commonly used to represent~\gls{RF} data and allow for logarithmic amplitude transformations. The rest of our network is built around the proposed time-frequency processing with the main workload to estimate two masks for separating the input mixture into two components being done in the Feature Transformer module.
\subsection{Preprocessing}
The mixed input signal in time domain $\textbf{x}_{\textit{m}}\in\mathbb{R}^{\textrm{1}\times{\textit{t}}}$ is transformed into its time-frequency representation $\textbf{X}_{\textit{m}}\in\mathbb{C}^{\textit{F}\times{\textit{T}}}$ using the discrete~\gls{STFT}~\cite{stft}
\begin{align}
	\mathbf{X}_{m} = \stft\left(\mathbf{x}_{m}\right),
	\label{eq:stft}
\end{align}
where $\textit{t}$ is the number of time steps of the time domain signal and $\textit{T}$ and $\textit{F}$ are the number of time and frequency bins of its~\gls{STFT}. The complex~\gls{STFT} is then converted into its exponential form and by applying a $\log_{10} $-transform to the absolute values, the potentially very large dynamic range of input amplitudes from different signals is squeezed:
\begin{align}
	&\mathbf{X}_{m\_logamp} = \log_{10}\left( \left\lvert \mathbf{X}_{m} \right\rvert \right)\\
	&\mathbf{X}_{m\_phase} = \atantwo\left(\Im\left(\mathbf{X}_m\right),\Re\left(\mathbf{X}_m\right)  \right).
\end{align}
 Through stacking the log-amplitude $\textbf{X}_ {\textit{m\_logamp}}\in\mathbb{R}^{\textit{F}\times{\textit{T}}}$ and phase information $\textbf{X}_ {\textit{m\_phase}}\in\mathbb{R}^{\textit{F}\times{\textit{T}}}$ along a new, third dimension, the 3D input tensor $\textbf{X}_ {\textit{m}2}\in\mathbb{R}^{\textit{F}\times{\textit{T}}\times{\textrm{2}}}$ for the Feature Extractor is formed.
\subsection{Feature Extractor}
The 3D input to the Feature Extractor is processed by a SeparableConv2D layer~\cite{dl_chollet, dl_goodfellow} followed by~\gls{LN}~\cite{layernorm}, building the intermediate Feature Extractor representation $\textbf{X}_{\textit{e}1}\in\mathbb{R}^{\textit{F}\times{\textit{T}}\times{\textit{N}}}$, where $N$ denotes the number of kernels in the convolution layer
\begin{align}
	\mathbf{X}_{e1}=\LN\left(\sconv2d\left(\mathbf{X}_{m2}\right)\right).
\end{align}
In the next step, another SeparableConv2D layer is applied to $\textbf{X}_{\textit{e}1}$, which allows for downsampling along the time and frequency axis by using $\textrm{strides}>\textrm{1}$. The convolution output is again normalized with~\gls{LN} and now nonlinearly activated with the~\gls{ReLU}~\cite{dl_goodfellow}. The second Feature Extractor signal $\textbf{X}_{\textit{e}2}\in\mathbb{R}^{\textit{F'}\times{\textit{T'}}\times{\textit{N}}}$ is therefore given by
\begin{align}
	\mathbf{X}_{e2}=\relu\left(\LN\left(\sconv2d\left(\mathbf{X}_{e1}\right)\right)\right)
\end{align}
with new dimensions $F'$ and $T'$, resulting from the strides in the convolution.
The output of the Feature Extractor $\textbf{X}_{e}\in\mathbb{R}^{\textit{F'}\times{\textit{T'}}\times{\textit{N}}}$ is built by passing $\textbf{X}_{e2}$ through a Conv2D layer~\cite{dl_chollet, dl_goodfellow} with subsequent~\gls{LN} since we realized, that adding another convolutional layer after downsampling improves accuracy:
\begin{align}
	\mathbf{X}_{e}=\LN\left(\nconv2d\left(\mathbf{X}_{e2}\right)\right).
\end{align}
\subsection{Feature Transformer}
\begin{figure}
	\centering
	\begin{subfigure}[b]{0.5\textwidth}
		\centering
		\includegraphics[width=\textwidth]{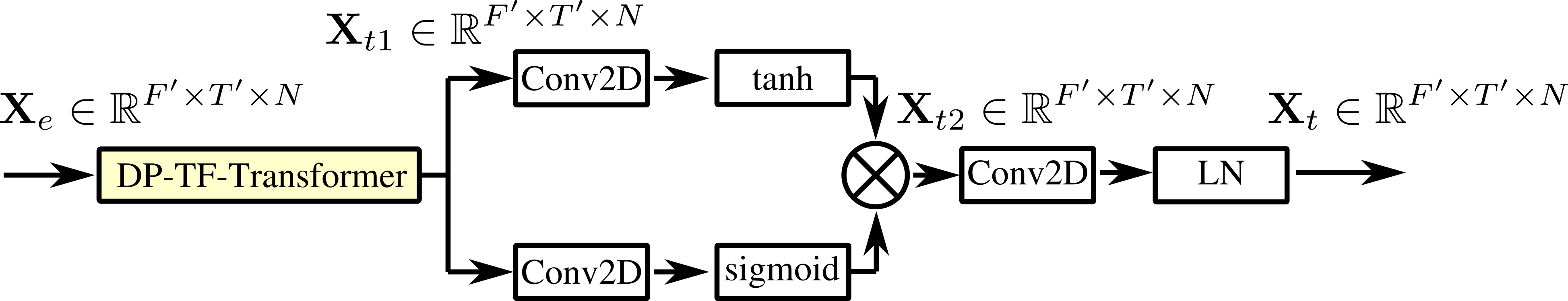}
		\caption{The Feature Transformer block with~\gls{DP-TF-Transformer} in yellow and gated output processing.}
		\label{fig:dptfta}
	\end{subfigure}
	\hfill
	\vfill
	\begin{subfigure}[b]{0.5\textwidth}
		\centering
		\includegraphics[width=\textwidth]{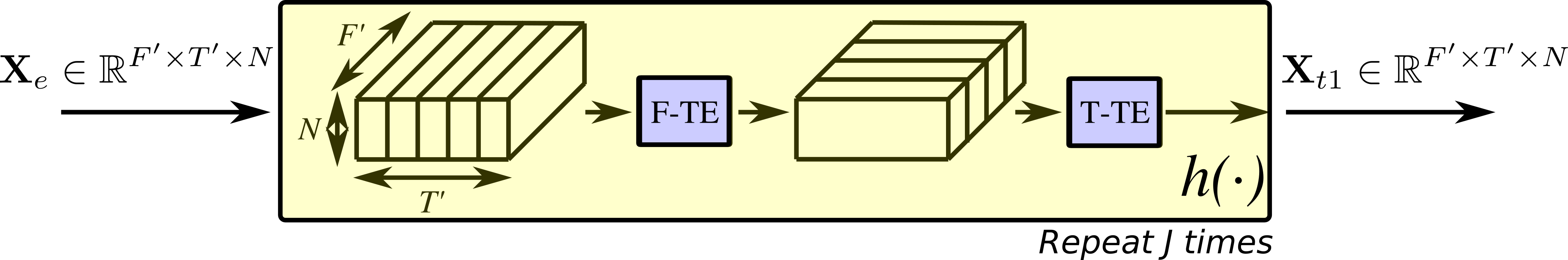}
		\caption{The~\gls{DP-TF-Transformer}, which consists of $J$ stacks of~\glspl{F-TE} and~\glspl{T-TE} in series.}
		\label{fig:dptftb}
	\end{subfigure}
	\hfill
	\vfill
	\begin{subfigure}[b]{0.5\textwidth}
		\centering
		\includegraphics[width=\textwidth]{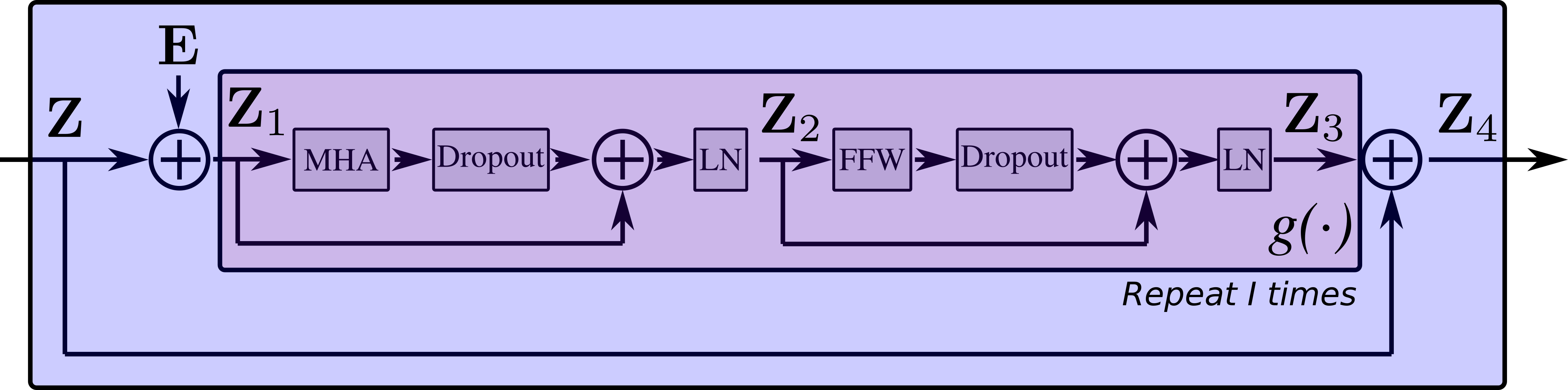}
		\caption{Structure of the Transformer Encoders that process one block in the~\glspl{F-TE} and~\glspl{T-TE}. First, the positional encoding is added. Then, $I$ stacks of vanilla Transformer Encoders follow. The input and output are connected by residual addition.}
		\label{fig:dptftc}
	\end{subfigure}
	\caption{Description of the Feature Transformer block.}
	\label{fig:separator}
\end{figure}

The Feature Transformer takes the Feature Extractor output and returns a block of the same shape. Its goal is to transform $\textbf{X}_{\textit{e}}\in\mathbb{R}^{\textit{F'}\times{\textit{T'}}\times{\textit{N}}}$ into a representation $\textbf{X}_{\textit{t}}\in\mathbb{R}^{\textit{F'}\times{\textit{T'}}\times{\textit{N}}}$ from which the mixed input can be separated effectively.
A high level view of the Feature Transformer is shown in Fig.~\ref{fig:dptfta}. The main component of the Feature Transformer and also of our entire model, which we call~\gls{DP-TF-Transformer}, is closely related to the SepFormer block in~\cite{trans_speech}, as it combines the parallelization and performance of Transformers~\cite{transformer} with the efficient long sequence modeling capabilities of dual-path processing introduced in~\cite{dprnn}.
The~\gls{DP-TF-Transformer} is described in Fig.~\ref{fig:dptftb}. It uses a stack of $\textit{J}$ modules, which are defined by the operation $h\left(\cdot\right)$. First comes a component we name~\gls{F-TE} that splits the input into T' frequency-channel-blocks of size $\textit{F'}\times{\textit{N}}$ and processes them separately. The~\gls{F-TE} applies a Transformer Encoder to each of these blocks. This Transformer Encoder attends to the frequencies and channels. The $\textit{T'}$ blocks which have passed through the~\gls{F-TE} are stacked together so that its input and output have equal shape. The same operation is then performed along the time instead of the frequency axis, where a~\gls{T-TE} processes $\textit{F'}$ time-channel-blocks of shape $\textit{T'}\times{\textit{N}}$ and concatenates all $\textit{F'}$ output blocks again. 
The mapping between~\gls{DP-TF-Transformer} input $\textbf{X}_{e}\in\mathbb{R}^{\textit{F'}\times{\textit{T'}}\times{\textit{N}}}$ and output $\textbf{X}_{t1}\in\mathbb{R}^{\textit{F'}\times{\textit{T'}}\times{\textit{N}}}$ can be described as
\begin{align}
	\mathbf{X}_{t1}=h^{J}\left(\mathbf{X}_{e}\right),
\end{align}
where $h^{J}\left(\cdot\right)$ denotes $\textit{J}$ applications of \gls{DP-TF-Transformer} modules.
The Transformer Encoders used to process one block are presented in detail in Fig.~\ref{fig:dptftc}. We stick to the vanilla Transformer implementation~\cite{transformer} and add sinusoidal positional encodings $\mathbf{E}\in\mathbb{R}^{\textit{F'}\times{\textit{N}}}$ or $\textbf{E}\in\mathbb{R}^{\textit{T'}\times{\textit{N}}}$ to the input $\textbf{Z}\in\mathbb{R}^{\textit{F'}\times{\textit{N}}}$ or $\textbf{Z}\in\mathbb{R}^{\textit{T'}\times{\textit{N}}}$ before each~\gls{F-TE} and~\gls{T-TE} block respectively
\begin{align}
	\mathbf{Z}_1=\textbf{Z}+\textbf{E}.
\end{align}
Such encoding is required since Transformers, as opposed to e.g.~\glspl{RNN}, have no intrinsic knowledge about the order of inputs. The encoded input of one block $\textbf{Z}_{1}$ is put through a number of $I$ Transformer Encoder layers, each implementing the operation $g \left(\cdot \right)$, as seen in the violet box of Fig.~\ref{fig:dptftc}. First, a~\gls{MHA} layer~\cite{transformer}, which performs scaled dot-product attention over the input with multiple heads, is employed. We then add Dropout~\cite{dl_goodfellow}, a residual addition around the~\gls{MHA} input and Dropout output and normalize with~\gls{LN}, resulting in $\textbf{Z}_2$:
\begin{align}
	\mathbf{Z}_2=\LN\left(\drop\left(\mha\left(\mathbf{Z}_1\right)\right)+\mathbf{Z}_1\right).
\end{align}
$\textbf{Z}_\textrm{2}$ is passed to a point-wise~\gls{FFW} network~\cite{transformer}, followed by Dropout, a residual addition around both operations and~\gls{LN}, which gives us $\textbf{Z}_\textrm{3}$ with
\begin{align}
	\mathbf{Z}_3=\LN\left(\drop\left(\ffw\left(\mathbf{Z}_2\right)\right)+\mathbf{Z}_2\right).
\end{align}
The mapping from $\textbf{Z}_\textrm{1}$ to $\textbf{Z}_\textrm{3}$ defined by the function $g\left(\cdot\right)$ is repeated $\mathit{I}$ times, before a residual addition around the whole block forms the output 
\begin{align}
	\mathbf{Z}_4=g^{I}\left(\mathbf{Z}_\mathrm{1}\right)+\mathbf{Z}.
\end{align}
Note that the stack of blocks in an~\gls{F-TE} and~\gls{T-TE} is processed in parallel, which speeds up the architecture. The Transformer Encoders that operate on the different blocks in one of $I$ layers also have shared weights, i.e. every block in such stack is processed with the same Transformer Encoder. This keeps the number of parameters and the computational complexity relatively low. At the head of our Feature Transformer, we apply a gating mechanism found in some audio models \cite{lafurca, dptnet} for 1D signals, that produces slight performance improvements. The gating output $\textbf{X}_{{\textit{t}2}}\in\mathbb{R}^{\textit{F'}\times{\textit{T'}}\times{\textit{N}}}$ is calculated as
\begin{equation}
	\begin{aligned}
		\mathbf{X}_{t2}=&\tanh\left(\nconv2d\left(\mathbf{X}_{t1}\right)\right)\circ \\
		&\mathrm{sigmoid}\left(\nconv2d\left(\mathbf{X}_{t1}\right)\right),
	\end{aligned}
\end{equation}
where $\circ$ denotes the Hadamard product~\cite{dl_goodfellow} and sigmoid and tanh~\cite{dl_goodfellow} are activation functions. The final Feature Transformer output $\textbf{X}_{\textit{t}}\in\mathbb{R}^{\textit{F'}\times{\textit{T'}}\times{\textit{N}}}$ then becomes
\begin{align}
	\mathbf{X}_\mathit{t}=\LN\left(\nconv2d\left(\mathbf{X}_{\mathit{t}\mathrm{2}}\right)\right).
\end{align}
\subsection{Separator}
In the Separator, the processed Feature Transformer output is first put through a Conv2DTranspose layer~\cite{t_conv} with the same stride as the downsampling SeparableConv2D layer in the Feature Extractor module. Thereby, the frequency and time dimensions are upsampled to their original shape. We then normalize with~\gls{LN} and activate this output with~\gls{ReLU}, resulting in the intermediate Separator signal $\textbf{X}_ {\textit{s}1}\in\mathbb{R}^{\textit{F}\times{\textit{T}}\times{\textit{N}}}$ with
\begin{align}
	\mathbf{X}_{s1}=\relu\left(\LN\left(\tconv2d\left(\mathbf{X_{\mathit{t}}}\right)\right)\right).
\end{align}
For the estimation of wanted filter masks $\textbf{X}_ {\textit{s}}\in\mathbb{R}^{\textit{F}\times{\textit{T}}\times{\textrm{4}}}$, we employ a Conv2D layer with 4 kernels and linear activation:
\begin{align}
	\mathbf{X}_{\mathit{s}}= \nconv2d\left(\mathbf{X}_{s1}\right).
\end{align} 
The first half of $\textbf{X}_{\textit{s}}$ are the log-amplitudes and the second half the phases of both masks.
\subsection{Postprocessing}
The log-amplitude and phase information from the Separator is then split. Log-amplitudes $\textbf{X}_ {\textit{s\_{logamp}}}\in\mathbb{R}^{\textit{F}\times{\textit{T}}\times{\textrm{2}}}$ are converted back to amplitudes in linear scale $\textbf{X}_ {\textit{s\_{linamp}}}\in\mathbb{R}^{\textit{F}\times{\textit{T}}\times{\textrm{2}}}$ using 
\begin{align}
	\mathbf{X}_{\mathit{s\_{linamp}}}= 10^{\mathbf{X}_ {\mathit{s\_{logamp}}}}.
\end{align}
With the amplitude and phase information $\textbf{X}_ {\textit{s\_{phase}}}\in\mathbb{R}^{\textit{F}\times{\textit{T}}\times{\textrm{2}}}$, we can then construct the complex valued filter masks $\textbf{X}_ {o}\in\mathbb{C}^{\textit{F}\times{\textit{T}}\times{\textrm{2}}}$ with
\begin{equation}
\begin{aligned}
	\mathbf{X}_{o} = \:&\mathbf{X}_{\mathit{s\_{linamp}}}\circ\cos(\mathbf{X}_ {\mathit{s\_{phase}}})+\\ \:&\mathbf{X}_{\mathit{s\_{linamp}}}\circ j\sin(\mathbf{X}_{\mathit{s\_{phase}}}).
\end{aligned}
\end{equation}
By splitting $\textbf{X}_ {\textit{o}}$ along the last dimension, we get the \gls{STFT} masks of the two signals $\textbf{X}_{\textit{o}1}\in\mathbb{C}^{\textit{F}\times{\textit{T}}}$ and $\textbf{X}_{\textit{o}2}\in\mathbb{C}^{\textit{F}\times{\textit{T}}}$. These are multiplied element-wise with the \gls{STFT} of the input mixture and the separated output signals in time domain $\textbf{x}_{\textit{o}1}\in\mathbb{R}^{\textrm{1}\times{\textit{t}}}$ and $\textbf{x}_{\textit{o}2}\in\mathbb{R}^{\textrm{1}\times{\textit{t}}}$ are formed by discrete \gls{iSTFT}:
\begin{align}
	&\mathbf{x}_{{o1}} = \istft \left(\mathbf{X}_{{o1}}\circ \mathbf{X}_{m}\right) \\
	&\mathbf{x}_{{o2}} = \istft \left(\mathbf{X}_{{o2}}\circ \mathbf{X}_{m}\right).
\end{align}
Since all operations of our network are differentiable, we can train the whole model end-to-end in time domain with backpropagation. Separating in time domain enables us to apply well established and powerful loss functions from the audio field, which circumvents the problem of finding suitable time-frequency losses.
\section{Experiments}
\subsection{Architecture Details}
For the~\gls{STFT} and~\gls{iSTFT}, we use a Hann window~\cite{stft} of size $\textrm{512}$ with a hop size of $\textrm{256}$ discrete time steps and therefore transform $\textbf{x}_{m}\in\mathbb{R}^{1\times{\textrm{65280}}}$ into its time-frequency representation $\textbf{X}_{m}\in\mathbb{C}^{\textrm{257}\times{\textrm{256}}}$, where we use only the positive frequency bins, which contain all information because $\textbf{x}_{m}$ is real-valued. The SeparableConv2D layers in the Feature Extractor and the Conv2DTranspose layer in the Separator have a kernel size of 4\texttimes4, all Conv2D layers use 1\texttimes1 kernels. We set 2\texttimes2 strides for down- and upsampling in the corresponding convolutional Feature Extractor and Separator layers, which allows for a bigger model at the cost of compressing the signal in frequency and time. To avoid information loss in the downsampling operation, we add one zero-padded frequency bin beforehand. Consequently, the Separator output has the dimension (\textit{F}+1)\texttimes \textit{T}\texttimes{4} and we delete the last frequency component again to be able to multiply estimated masks with the mixed input. For simplicity, the feature dimension $\textit{N}$ is kept constant throughout the model. This includes the model- and inner feed-forward dimension of the Transformer Encoders and all convolutional layers except for the Separator output. We set $\textit{N}=\textrm{128}$ heuristically, such that the available~\gls{GPU} memory of 8 GB (Nvidia GeForce GTX 1080) is maxed out while training with a batch size of $\textrm{2}$. \gls{MHA} layers have 2 attention heads and we choose $\textit{I}=\textit{J}=\textrm{2}$ for the \gls{DP-TF-Transformer} with a dropout rate of $0.1$ in the Dropout layers. For positional encoding, we stick to the original settings in~\cite{transformer}. Our model has $1.145.764$ parameters in total.
\subsection{Dataset}
\begin{figure}
	\centering{
	\includegraphics[width=1.0\linewidth]{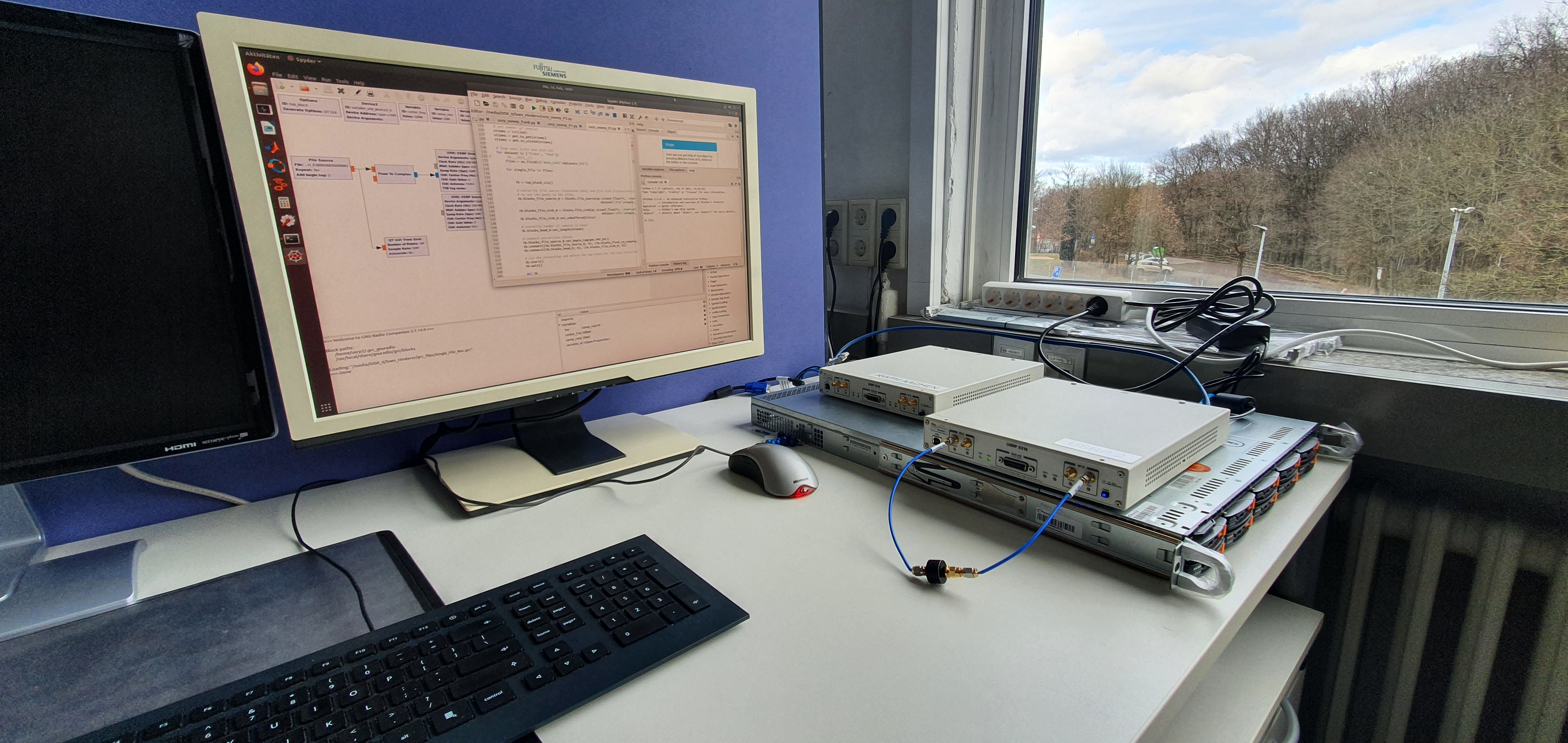}}
	\caption{Simple setup for real data generation: Test samples are automatically transmitted and received by a \gls{SDR} (USRP X310 by Ettus) as seen on the right. The channel is a cable with a 6\,dB attenuator.}
	\label{fig:messungsdr}
\end{figure}

Datasets are created by simulating common~\gls{LPI} waveforms using GNU Radio~\cite{gnuradio} and Python. To evaluate the algorithm with both simulated and real signals, we transmit and receive all test samples with a~\gls{SDR} in a simplistic setup, shown in Fig.~\ref{fig:messungsdr}. As intrapulse modulation, we choose P1, Costas and Frank codes~\cite{Levanon_2004_RadarSignals} for training and P3, Barker and linear chirps~\cite{Levanon_2004_RadarSignals} for testing. Interpulse modulations can be either~\gls{CW}, constant~\gls{PRI}, stagger~\gls{PRI}, jitter~\gls{PRI} or Dwell and Switch. Due to hardware limitations, a sample rate of 50~MHz is used and all signals are in the frequency range between 0~MHz and 25~MHz, an extension to shift this to an arbitrary band would however be straight forward. \newline
For each intrapulse modulation type, we generate 2000 samples for training and 1000 for testing with a length of 10\textsuperscript{6} time steps. To increase the variance of the dataset, which is necessary to achieve a high level of generalization of the trained neural network, we randomly choose a new set of parameters each sample. These are described in Table~\ref{tab_pulse_parameters}.  
\begin{table}[htbp]
\caption{Intrapulse parameters of the used waveforms. Samples are drawn uniformly from given values or value ranges.}
\setlength\arrayrulewidth{1.pt}
\begin{center}
\begin{tabular}{lll}
	\textbf{Intrapulse modulation} & \textbf{Parameters} & \textbf{Values}           \\ 
	\hline
	\multirow{2}{*}{All}           & PW                  & [4, 50)\,\textmu s                \\
	& PRI                 & [2PW,~2PW + [50, 400)\,\textmu s)          \\ 
	\hline
	Costas                         & bandwidth           & [5, 23) MHz               \\
	& code length         & $\left\{3,4,5,6,8,9,10\right\}$       \\
	Frank                          & frequency           & [3, 23)\,MHz        \\
	& code length         & [3, 8]                     \\
	P1                             & frequency           & [3, 23)\,MHz                \\
	& code length         & [3, 8]                       \\ 
	\hline
	Linear chirp                   & start frequency     & [3, 15)\,MHz           \\
	& bandwidth           & [2, 23 - start\_freq)\,MHz  \\
	Barker                         & frequency           & [3, 23)\,MHz                \\
	& code length         & $\left\{2,3,4,5,7,11,13\right\}$       \\
	P3                             & frequency           & [3, 23)\,MHz              \\
	& code length         & [3, 20] \\ \hline                   
\end{tabular}
\label{tab_pulse_parameters}
\end{center}
\end{table}
\subsection{Training Procedure}
Every epoch, a new training set is created to further increase our database artificially. The steps are as follows:
\begin{enumerate}
	\item Randomly draw 3000 pairs of training signals to be mixed from the 6000 simulated waveforms.
	\item Cut out chunks of size 65280, i.e. the input and output signal lengths of the network, from the saved signals with 10\textsuperscript{6} time steps. The position is chosen at random.
	\item Scale the signals such that they have amplitudes in the range of -10 dBm to -80 dBm. 
	\item Corrupt them by adding Gaussian noise of various levels.
	\item Mix the signals by adding them together.
	\item Normalize the mixture and ground truth signals by dividing their amplitudes by the maximum amplitude of the input mixture.
\end{enumerate}
Our test set is built with the same procedure, utilizing all independent and real test samples once.
We use the~\gls{SD-SDR}~\cite{sd-sdr} as training objective, which is defined as
\begin{align}
\textrm{SD-SDR}\left(s,\hat{s}\right)=10~\mathrm{log_{10}}\left(\frac{ \left\Vert\frac{\hat{s}^{T}s}{ \left\Vert s \right\Vert^2 } s\right\Vert^2}{ \left\Vert s-\hat{s} \right\Vert^2 }\right),
\end{align}
where $\textit{s}$ is the "ground truth" signal in the machine learning sense and $\hat{\textit{s}}$ its estimate. \gls{SD-SDR} has shown to provide good separation results while preserving important amplitude information. Utilizing~\gls{uPIT}~\cite{upit,nachmani2020voice} to tackle the output permutation problem that is associated with blind source separation, we get the following loss function $\textit{l}\left(\textit{s},\hat{\textit{s}}\right)$
\begin{align}
	l\left(s,\hat{s}\right)=-\max_{\pi\in\Pi_C}\frac{ 1 }{ C }\sum_{ c=1 }^{ C }{\textrm{SD-SDR}\left( s_{c} ,\hat{s}_{\pi\left( c \right)}\right)}.
\end{align}
$\Pi_\textit{C}$ is the set of possible output channel permutations with $\textit{C}$ being the number of channels. This function optimizes the average SD-SDR of both signals. Output channels are permuted such that this objective is maximized. Permutation invariance is necessary, since we have no knowledge about the input waveforms and therefore can't assign certain signals, e.g. chirps, to a fixed network output channel.\newline
We train the models in TensorFlow~\cite{tensorflow2015-whitepaper} for 70 epochs, using the adam optimizer~\cite{dl_goodfellow} with default settings and a decaying learning rate $\textit{lr}$ defined by the schedule 
\begin{equation}
	\mathit{lr} = 10^{ -4 }*0.90^{\mathit{epoch}}.
\label{eq:learning_rate}
\end{equation}
We also apply the built-in loss scaling, which dynamically scales the loss after a forward pass. This is necessary to solve the problem of rare underflows during backpropagation, caused by the large value range of different signal amplitudes, even when running our network with 32 bit floating point and 64 bit complex precision. 
\section{Results}
\begin{figure}
	\centering{
		\includegraphics[width=1.0\linewidth]{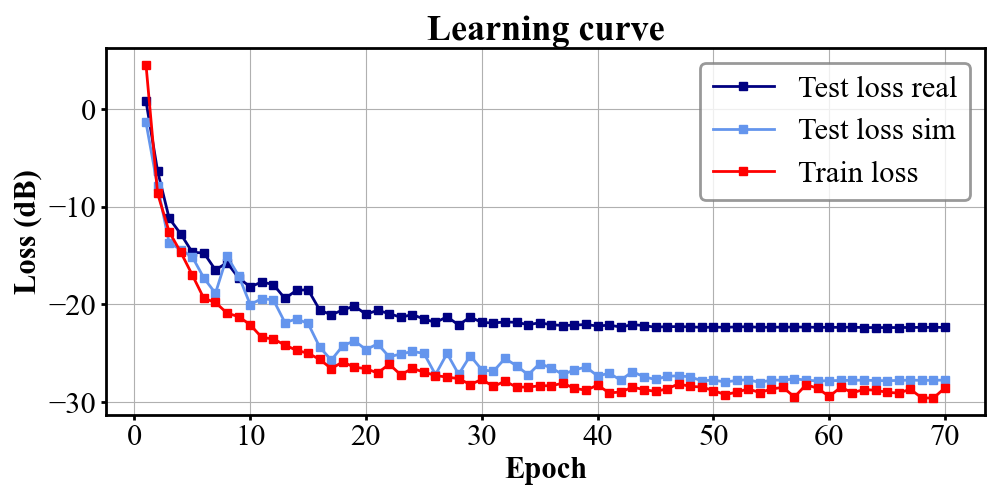}}
	\caption{Train and test losses with real and simulated data evaluated each training epoch.}
	\label{fig:training_curve}
\end{figure}

\begin{figure}
	\centering
	\begin{subfigure}[b]{.5\textwidth}\
		\includegraphics[width=\textwidth]{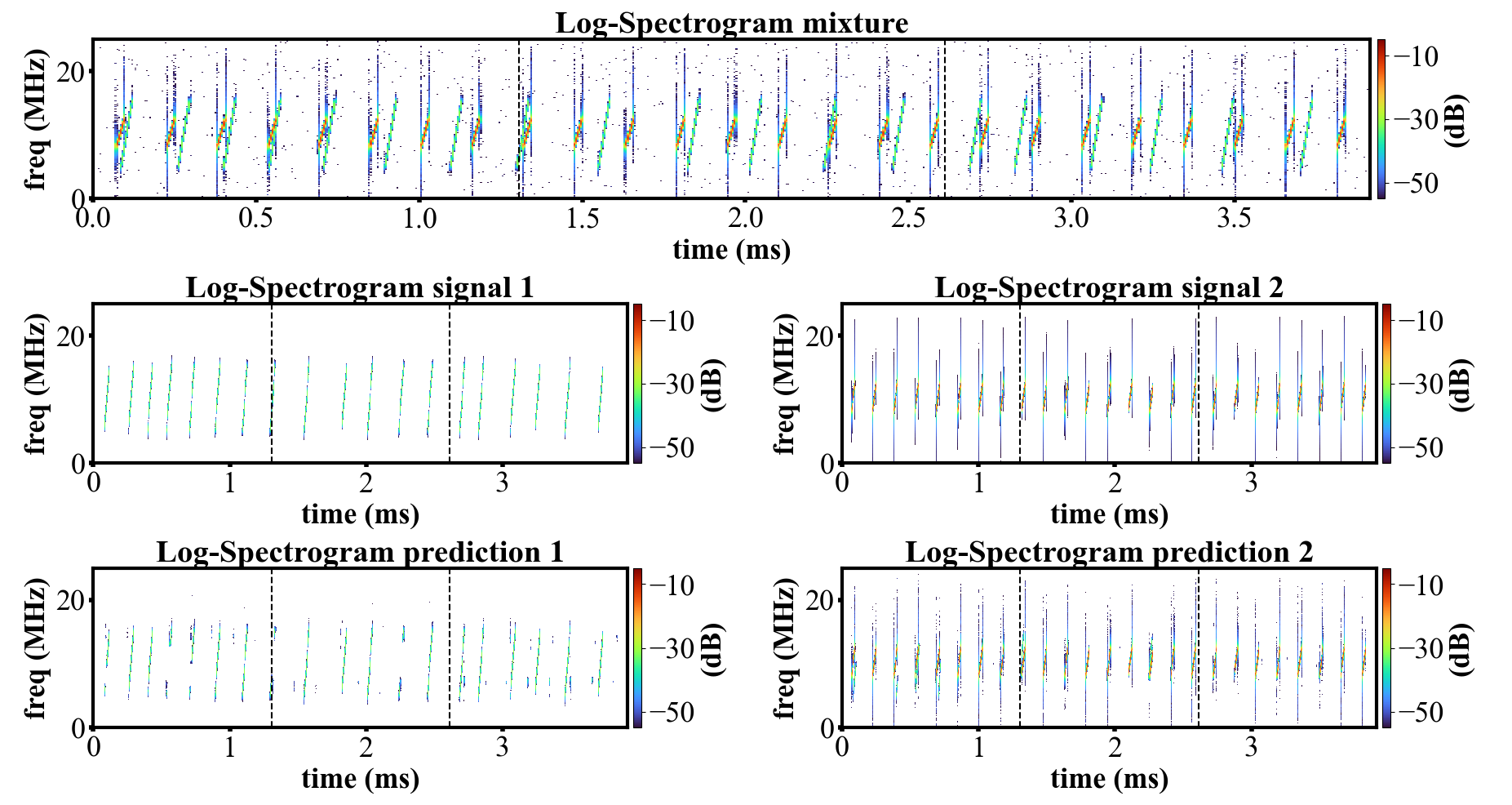}
		\caption{Similar chirps.}
		\label{fig:sepa}
	\end{subfigure}
	\hfill
	\begin{subfigure}[b]{0.5\textwidth}
		\centering
		\includegraphics[width=\textwidth]{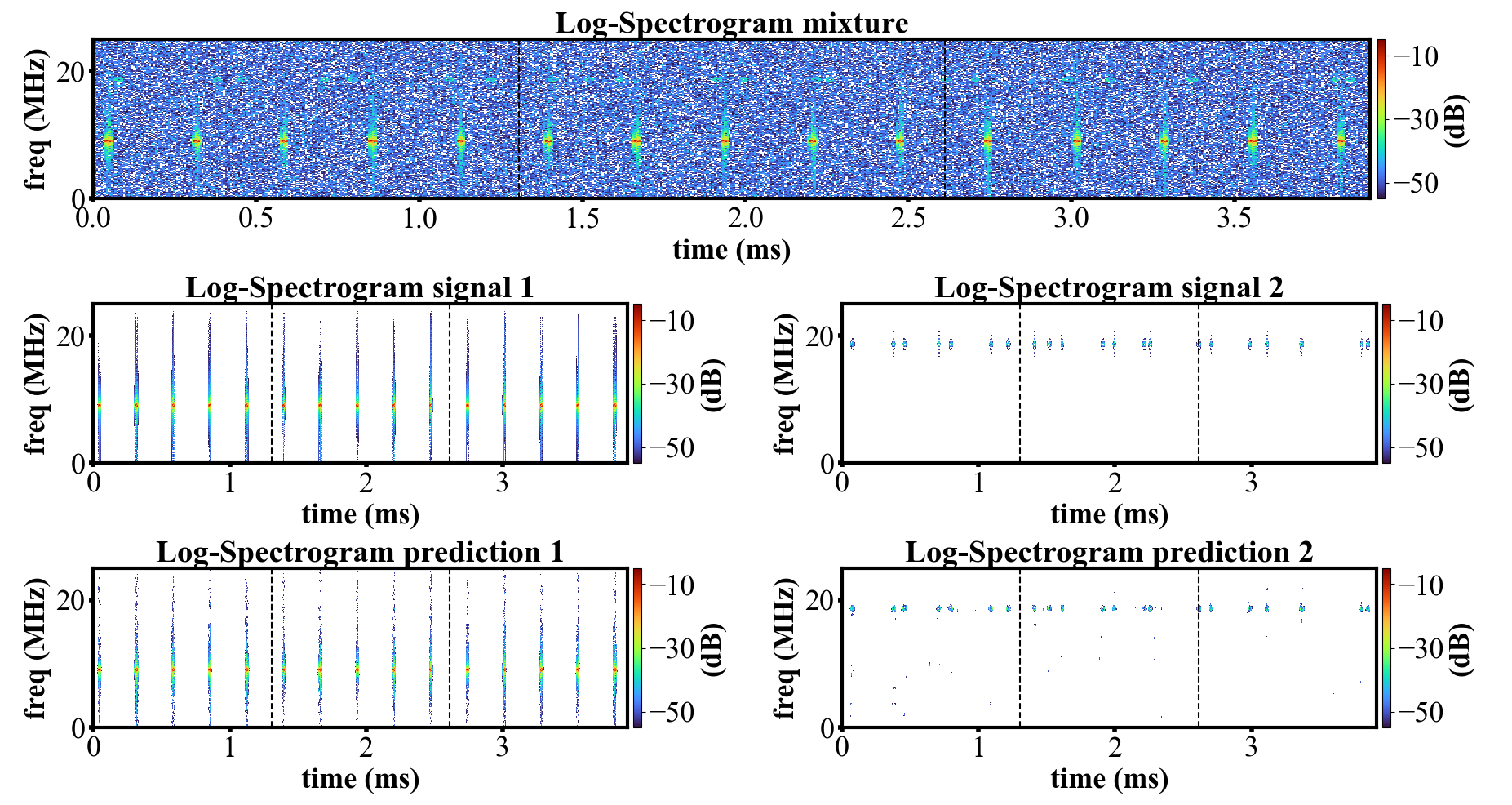}
		\caption{Low~\gls{SNR}.}
		\label{fig:sepb}
	\end{subfigure}
	\hfill
	\begin{subfigure}[b]{0.5\textwidth}
		\centering
		\includegraphics[width=\textwidth]{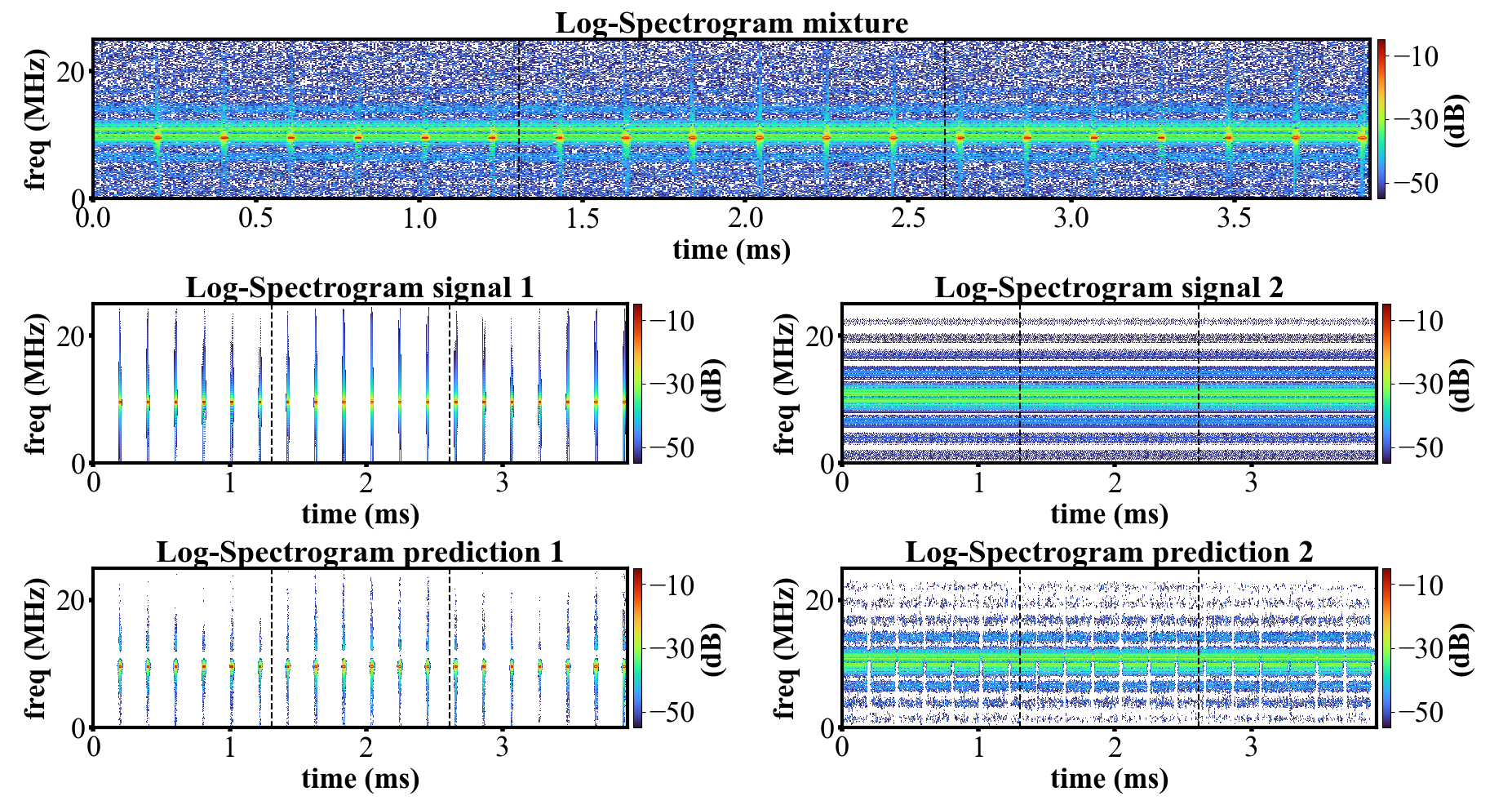}
		\caption{Pulsed and~\gls{CW} signals.}
		\label{fig:sepc}
	\end{subfigure}
	\hfill
\begin{subfigure}[b]{0.5\textwidth}
	\centering
	\includegraphics[width=\textwidth]{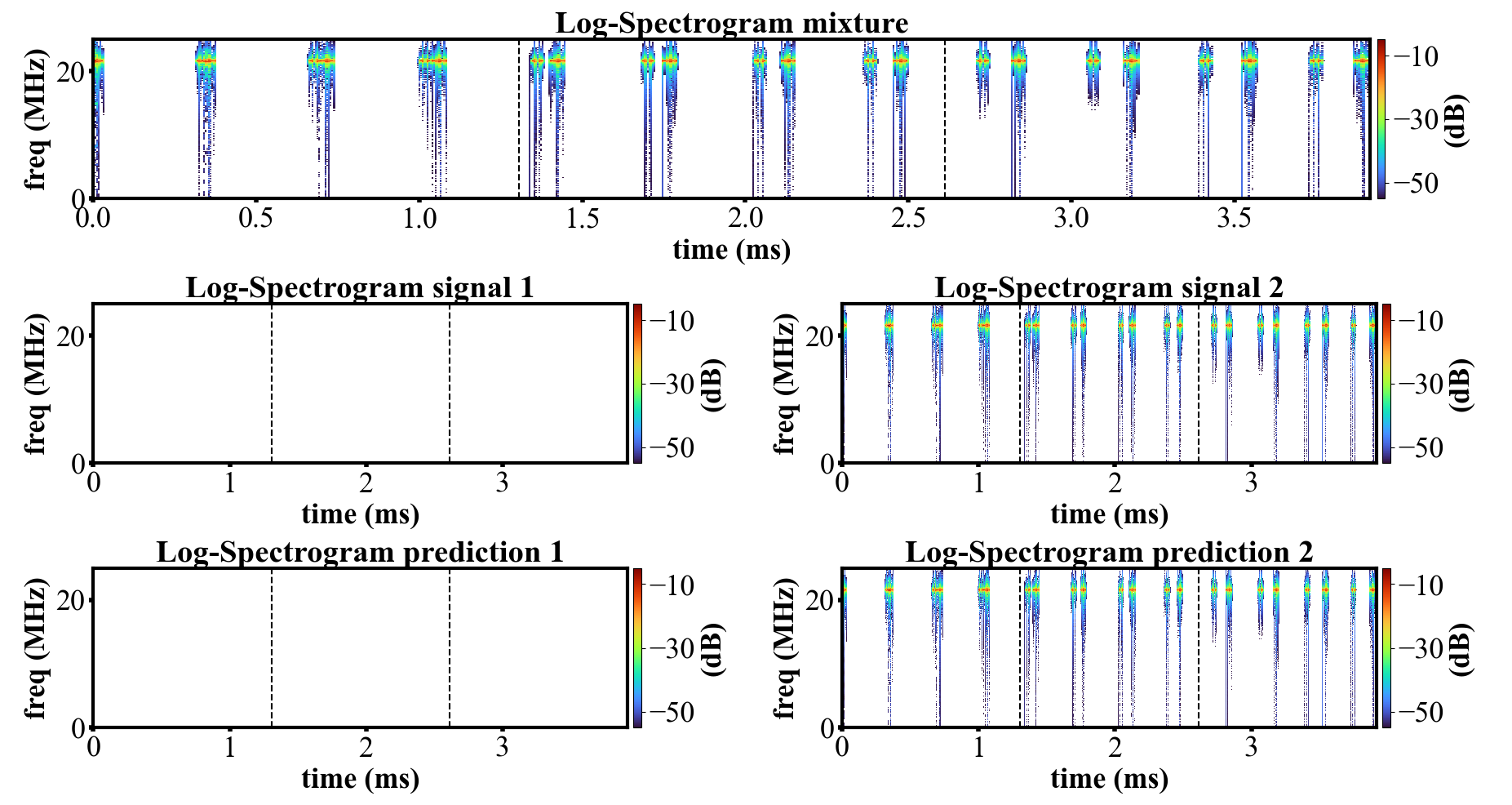}
	\caption{Only one signal.}
	\label{fig:sepd}
\end{subfigure}
	\caption{Separation under challenging conditions.}
	\label{fig:sep}
\end{figure}

The learning curve of the proposed model can be seen in Fig.~\ref{fig:training_curve}. 
It shows a clear performance enhancement over the course of training and converges smoothly. Train and test losses with simulated data are almost equal and the discrepancy between test losses using real and simulated data is expected since real test samples have lower~\gls{SNR} because of attenuation and other hardware effects. The seen behavior can be a result of underfitting\footnote{Subsequent studies should investigate this by training larger models.}. If this assumption is correct, a further enlargement of the model size, e.g. by increasing the feature dimension $\textit{N}$, could lead to better results. It also suggests that we are experiencing a memory bottleneck, mostly driven by the memory complexity of vanilla Transformer Encoders. \newline
Next, we demonstrate the capabilities of our system through presenting multiple separated test samples in Fig.~\ref{fig:sep}. We stack three observation periods together, each having the length of our model input and outputs. This is done to check for unwanted channel swaps that might occur since we train on restricted input lengths, i.e. one utterance in~\gls{uPIT}, between which output channels can change. 
Results show, that in none of the examples, channel swaps occur. They happen very rarely in the whole test set, which would require some additional matching or tracking to resolve the issue. Our model is able to separate similar chirps, see Fig.~\ref{fig:sepa}. Remember, that it has not seen any chirp during training. The closest and only frequency modulated waveforms in the training set are Costas codes, which look quite different. Our system also works under low~\gls{SNR} conditions (Fig.~\ref{fig:sepb}), can separate overlapping~\gls{CW} and pulsed waveforms (Fig.~\ref{fig:sepc}) and even functions if only one signal is present (Fig.~\ref{fig:sepd}), which has never been the case during training. Further note the implicit denoising, since the model was trained to predict clean and purely simulated data. Up to some examples where separation fails and which have to be studied in more detail in the future, the given examples are representative and proof the potential of the proposed method, especially given the fact that our training data contains only three different itrapulse modulations.
\section*{Conclusion}
We proposed a new algorithm for the separation of~\gls{RF} signals to overcome the limitations of current deinterleaving methods, which often require the signals to be distinguishable in frequency or~\gls{AOA}. Therefore, a neural network architecture has been developed that separates mixed signals in time domain. We have shown that our model is capable of separating two unknown waveforms with a single input channel and in a defined frequency band under difficult conditions. This might enable us to deal with scenarios, where current methods fail. The network architecture could also be used as a baseline for different tasks with~\gls{RF} signals or modified to separate signals in other domains. For future work, we see multiple research directions. To improve the given model, extending the simulation framework and collecting more labeled training data would likely be most effective. We have shown that memory complexity might be an issue. This becomes even more relevant, if larger bandwidths are to be investigated. Network pruning but especially memory efficient Transformer Encoders could alleviate this problem. Cases where separation fails should be studied in subsequent work. Extending the algorithm to handle an unknown number of signals would make the proposed system applicable to a broader range of problems. Finally, the performance of the system could be compared to classical methods if angular information is available by using multi channel receivers.
\section*{Acknowledgment}
We would like to thank Angel Slavov for his help with the real data generation.
\bibliographystyle{ieeetr}
\bibliography{ref}
\end{document}